\documentclass[twocolumn,showpacs,preprintnumbers,prb,aps,amssymb]{revtex4}
\usepackage{graphicx}
\usepackage{dcolumn}
\usepackage{bm}

\begin{document}

\preprint{preprint}

\title{Elastic Tensor of Sr$_2$RuO$_4$}

\author{Johnpierre Paglione, C. Lupien, W. A. MacFarlane, J. M. Perz and Louis Taillefer}

\affiliation{Department of Physics, University of Toronto, Toronto, Ontario M5S 1A7, Canada}

\author{Z. Q. Mao and Y. Maeno}

\affiliation{Department of Physics, Graduate School of Science, Kyoto University, Kyoto 606-8502, Japan}

\date{\today}

\begin{abstract} 

The six independent elastic constants of Sr$_2$RuO$_4$ were determined using resonant
ultrasound spectroscopy on a high-quality single-crystal specimen.  The constants are
in excellent agreement with those obtained from pulse-echo experiments performed on a
sample cut from the same ingot.  A calculation of the Debye temperature using the
measured constants agrees well with values obtained from both specific heat and
M\"{o}ssbauer measurements.

\end{abstract}


\maketitle


Since the discovery of superconductivity in Sr$_2$RuO$_4$,\cite{Maeno_Nature} there
has been much interest in the possibility of novel pairing symmetry in this
material.\cite{Maeno_PT} The availability of large, high-quality single crystals of
Sr$_2$RuO$_4$ has allowed the full range of physical properties to be measured, with
the prospect of identifying the order parameter symmetry and the mechanism of
superconductivity. A thorough knowledge of the elastic properties is one important
element in obtaining a complete understanding.

The elastic constants of a material, which relate deformation to stress, are of
interest because they are involved in fundamental solid-state phenomena: interatomic
potentials, equations of state and phonon spectra.  Furthermore, thermodynamics links
elasticity directly to quantities such as thermal expansivity, atomic volume,
Debye temperature, and Gr\"{u}neisen parameter.\cite{Truell} A determination of the
Debye temperature provides information on the phonon contribution to the
low-temperature specific heat and the possible role of electron-phonon coupling in
superconductivity.

There are six independent second-order elastic constants $C_{ij}$ associated with the
tetragonal crystal structure of Sr$_2$RuO$_4$. Using Voigt notation, they are
expressed as $C_{11}$, $C_{33}$, $C_{23}$, $C_{12}$, $C_{44}$, and $C_{66}$.  The
constants for which $i=j$ correspond to sound propagation in various principle
crystal directions.  When using a conventional time-of-flight, or pulse-echo,
measurement technique, the determination of off-diagonal constants requires the
troublesome measurement of sound propagation along non-principal directions,
something which has not been done for Sr$_2$RuO$_4$. This difficulty does not arise
when using resonance spectroscopy, which relies on a different technique to obtain
the full elastic tensor of a material in a single measurement.  In this article, we
report the full elastic tensor of Sr$_2$RuO$_4$ obtained by resonance spectroscopy.


The single-crystal sample of Sr$_2$RuO$_4$ was grown by the traveling solvent
floating zone technique \cite{Mao} and found to have a superconducting transition
temperature of 1.37 K as determined from magnetic susceptibility, which is a good
indication of its high purity and quality.  The sample was aligned using Laue x-ray
diffraction, cut into a rectangular parallelepiped using spark erosion, and polished
to dimensions of $2.88(1) \times 1.99(1) \times 0.86(1)$~mm$^3$. The sample mass was
$0.0292$~g, yielding a density of $5.92(8)~$g/cm$^3$, which agrees with the
previously reported value of $5.918~$g/cm$^3$.$~$\cite{density}

Measurements were made at room temperature using resonant ultrasound spectroscopy
(RUS).\cite{Migliori} This technique abandons the plane-wave approximation used in
pulse-echo experiments and instead uses the normal modes of vibration of a solid
specimen of known geometry, crystal symmetry and density to deduce the complete
elastic tensor in a single measurement.  Thus, it allows for relatively easy
determination of off-diagonal elastic constants, which are generally impractical to
measure in non-cubic symmetries using the pulse-echo technique.

The acoustic resonance frequencies of a single-crystal specimen can be calculated
given the dimensions, mass, and elastic constants.  The key to the RUS technique lies
in the ability to determine unknown parameters (in this case the elastic constants)
from knowledge of the resonance frequencies, which are readily determined
experimentally. There exists no analytic method of performing this calculation, and
so the unknown parameters are determined via a computational fitting procedure which
uses an iterative algorithm to match resonance frequencies calculated analytically
with those measured experimentally.  This calculation requires the input of measured
dimensions, mass, and an estimated set of elastic constants to start the fitting
iteration: the elastic constants are left as adjustable parameters to be determined
by minimization of error between the measured and calculated frequencies.  Because
the largest source of error in this experiment was the determination of the sample
dimensions, they were also left as free parameters in the fitting routine, allowing
for a comparison between measured dimensions and those determined from the fit. As a
test, the RUS apparatus and method were used to obtain the complete elastic tensor of
the heavy-fermion material UPt$_3$ (which has a hexagonal crystal structure). This
gave the values $C_{12}=1.44(2)$, $C_{23}=1.70(2)$, $C_{33}=2.93(2)$,
$C_{44}=0.3794(1)$, and $C_{66}=0.8367(9)$ (all in units of $10^{12}~$dynes/cm$^2$),
\cite{JP} which are in excellent agreement with those measured by de~Visser {\it et
al.} using the pulse-echo technique.\cite{DeVisser}

\begin{figure}
 \centering
 \includegraphics[totalheight=3in]{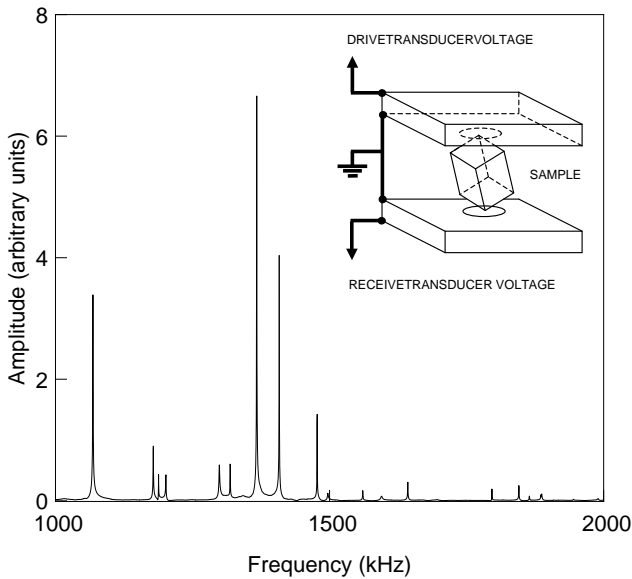}
 \caption{\label{fig:spectrum} A portion of the resonance spectrum of a rectangular parallelepiped of Sr$_2$RuO$_4$ obtained using the RUS technique. The inset shows a schematic of the RUS apparatus.}
\end{figure}

The resonance frequencies of the Sr$_2$RuO$_4$ sample were determined at room
temperature using the apparatus shown schematically in the inset of
Fig.~\ref{fig:spectrum}.  By holding the sample between two piezoelectric
transducers, where one transducer is used to drive the sample with mechanical
vibrations and the other to detect the mechanical response of the sample, a frequency
sweep allowed a rapid measurement of the resonance modes. It should be noted that the
only assumption made in the RUS technique lies in the ability to observe the natural
resonances of a solid body under \textit{free} boundary conditions, allowing a proper
comparison between measurement and calculation.  The rectangular parallelepiped
sample was held only lightly by its corners to minimize suppression of any resonance
modes, thus making the free boundary approximation appropriate.

Measurements up to $\sim3$~MHz provided more than sixty measured resonance
frequencies. The central portion of this frequency spectrum is shown in
Fig.~\ref{fig:spectrum}. The first thirty measured frequencies are listed in
Table~\ref{tab:modes} along with those calculated analytically using a program
developed specifically for the rectangular parallelepiped sample
geometry.\cite{Migliori} Note that all of the first thirty calculated resonance
frequencies were detected, and only three of the next thirty were undetectable. The
error is listed as the percentage difference between measured and calculated values.
It should be noted that in RUS measurements it is commonly found that the first one
or two modes typically have a much larger discrepancy between measured and calculated
values than the average, as is seen for the first mode in Table~\ref{tab:modes}, and
are therefore not weighted in the fit.  The dependence of each mode on the elastic
constants was also estimated using this program in order to indicate the pure
(shear/compressional) or composite nature of each resonance mode, and are listed as
the normalized values $df/dC_{ij}$ in Table~\ref{tab:modes}. For example, the $n=5$
resonance mode can be seen to depend mainly on the constant $C_{66}$, and is thus
primarily a pure shear mode.  Identification of pure resonance modes can be of use in
tracking the behavior of individual elastic constants as a function of temperature or
magnetic field.

\begin{table*}
\caption{\label{tab:modes} Comparison between measured and calculated values of the first 30 resonance frequencies of the Sr$_2$RuO$_4$ sample. Included are the normalized $df/dC_{ij}$ values, which indicate the nature of each resonance mode $n$.}
\begin{ruledtabular}
\begin{tabular}{cccccccccc}
 $n$ & $f_{meas}$ (MHz) & $f_{calc}$ (MHz) & \textbar~error \%~\textbar & $df/dC_{11}$ & $df/dC_{33}$ & $df/dC_{23}$ & $df/dC_{12}$ & $df/dC_{44}$ & $df/dC_{66}$\\
\hline
  1& 0.379947& 0.373526& 1.69\footnotemark[1]&   0.01&  0.00&  0.00&  0.00&  0.21&  0.78\\
  2& 0.461821& 0.461405& 0.09&   1.32&  0.03& -0.07& -0.43&  0.15&  0.00\\
  3& 0.698196& 0.697239& 0.14&   1.09&  0.04& -0.08& -0.32&  0.00&  0.27\\
  4& 0.761808& 0.760115& 0.22&   0.34&  0.01& -0.02& -0.11&  0.30&  0.48\\
  5& 0.855415& 0.853657& 0.21&   0.10&  0.01& -0.01& -0.02&  0.00&  0.93\\
  6& 0.868241& 0.868475& 0.03&   0.92&  0.07& -0.13& -0.11&  0.25&  0.01\\
  7& 0.902003& 0.902758& 0.08&   1.70&  0.02& -0.04& -0.68&  0.00&  0.00\\
  8& 0.907763& 0.908625& 0.09&   1.06&  0.01& -0.01& -0.44&  0.29&  0.10\\
  9& 1.069506& 1.068929& 0.05&   0.53&  0.06& -0.12&  0.04&  0.42&  0.07\\
 10& 1.179531& 1.179844& 0.03&   0.57&  0.01& -0.02& -0.21&  0.43&  0.21\\
 11& 1.187974& 1.186127& 0.16&   1.12&  0.05& -0.08& -0.35&  0.00&  0.27\\
 12& 1.200415& 1.198150& 0.19&   1.43&  0.06& -0.11& -0.44&  0.00&  0.06\\
 13& 1.299958& 1.301070& 0.09&   0.42&  0.02& -0.04& -0.09&  0.39&  0.29\\
 14& 1.318483& 1.317963& 0.04&   0.60&  0.02& -0.03& -0.22&  0.00&  0.63\\
 15& 1.366880& 1.366446& 0.03&   1.50&  0.01& -0.02& -0.64&  0.00&  0.15\\
 16& 1.408010& 1.408700& 0.05&   0.48&  0.03& -0.05& -0.12&  0.00&  0.67\\
 17& 1.477300& 1.480136& 0.19&   1.12&  0.15& -0.25& -0.05&  0.00&  0.03\\
 18& 1.498835& 1.496335& 0.17&   0.62&  0.01& -0.02& -0.23&  0.58&  0.04\\
 19& 1.561982& 1.561620& 0.02&   0.56&  0.03& -0.06& -0.10&  0.55&  0.02\\
 20& 1.596411& 1.592585& 0.24&   0.38&  0.02& -0.03& -0.08&  0.66&  0.05\\
 21& 1.643878& 1.639129& 0.29&   0.36&  0.03& -0.06& -0.02&  0.64&  0.06\\
 22& 1.697681& 1.702562& 0.29&   0.26&  0.02& -0.03& -0.04&  0.47&  0.31\\
 23& 1.796324& 1.797892& 0.09&   1.05&  0.07& -0.10& -0.31&  0.00&  0.29\\
 24& 1.845635& 1.846354& 0.04&   0.65&  0.03& -0.04& -0.23&  0.00&  0.59\\
 25& 1.864200& 1.862988& 0.07&   0.66&  0.04& -0.06& -0.20&  0.00&  0.56\\
 26& 1.885902& 1.887157& 0.07&   1.53&  0.02& -0.03& -0.65&  0.00&  0.13\\
 27& 1.887384& 1.889048& 0.09&   0.29&  0.02& -0.04& -0.01&  0.62&  0.13\\
 28& 1.908689& 1.900802& 0.41&   0.23&  0.01& -0.01& -0.07&  0.83&  0.01\\
 29& 1.945637& 1.942492& 0.16&   0.21&  0.01& -0.03& -0.03&  0.71&  0.12\\
 30& 1.992652& 1.995493& 0.14&   0.43&  0.02& -0.04& -0.11&  0.57&  0.13\\
\end{tabular}
\end{ruledtabular}
\footnotetext[1]{This mode was not weighted in the fit, as explained in the text.}
\end{table*}


The elastic constants used as initial fitting parameters were estimated using a
combination of available values for Sr$_2$RuO$_4$\cite{Matsui67,Lupien} and the
isostructural material La$_2$CuO$_4$\cite{Sarrao}, and subsequently adjusted
through numerous fitting iterations to obtain the best fit. The quality of each
fit was calculated using the root-mean-square error $\sigma_{RMS}$ between
measured and calculated resonant frequencies,
\begin{equation}
 \sigma_{RMS} = 100 \times \sqrt{ \frac{1}{N} \sum_{n=1}^N \bigg(
\frac{f_{meas}^{(n)}-f_{calc}^{(n)}}{f_{calc}^{(n)}} \bigg)^2} ~~\%,
\end{equation} 
where $n$ is the mode number and $N$ is the total number of modes weighted in the
fit. The fitting procedure resulted in a $\sigma_{RMS}$ value of $0.19\%$ between
the measured and calculated frequencies: such a small value is an indication of an
excellent fit.\cite{Migliori} The fitted dimensions agreed with the measured
values to within error, also confirming the validity of the fit. The elastic
constants calculated using this fit are compared in Table~\ref{tab:moduli} with
values obtained for Sr$_2$RuO$_4$ from various pulse-echo experiments and with
those obtained for La$_2$CuO$_4$ using the RUS technique. The absolute accuracy of
the constants obtained in this study was based on either a quality-of-fit
estimate, obtained from a measure of the stability of the minimization calculation
used in the fitting routine, or on errors in density measurement, taking the
larger of the two as the conservative approximation.  Specifically, the errors on
$C_{11}$, $C_{33}$, $C_{23}$, and $C_{12}$ were obtained from quality-of-fit
estimates, and those on $C_{44}$ and $C_{66}$ were based on the variation in
calculated $C_{ij}$ values obtained from fits using upper and lower dimensional
bounds as input parameters.\cite{error}

\begin{figure}
 \centering
 \includegraphics[totalheight=3in]{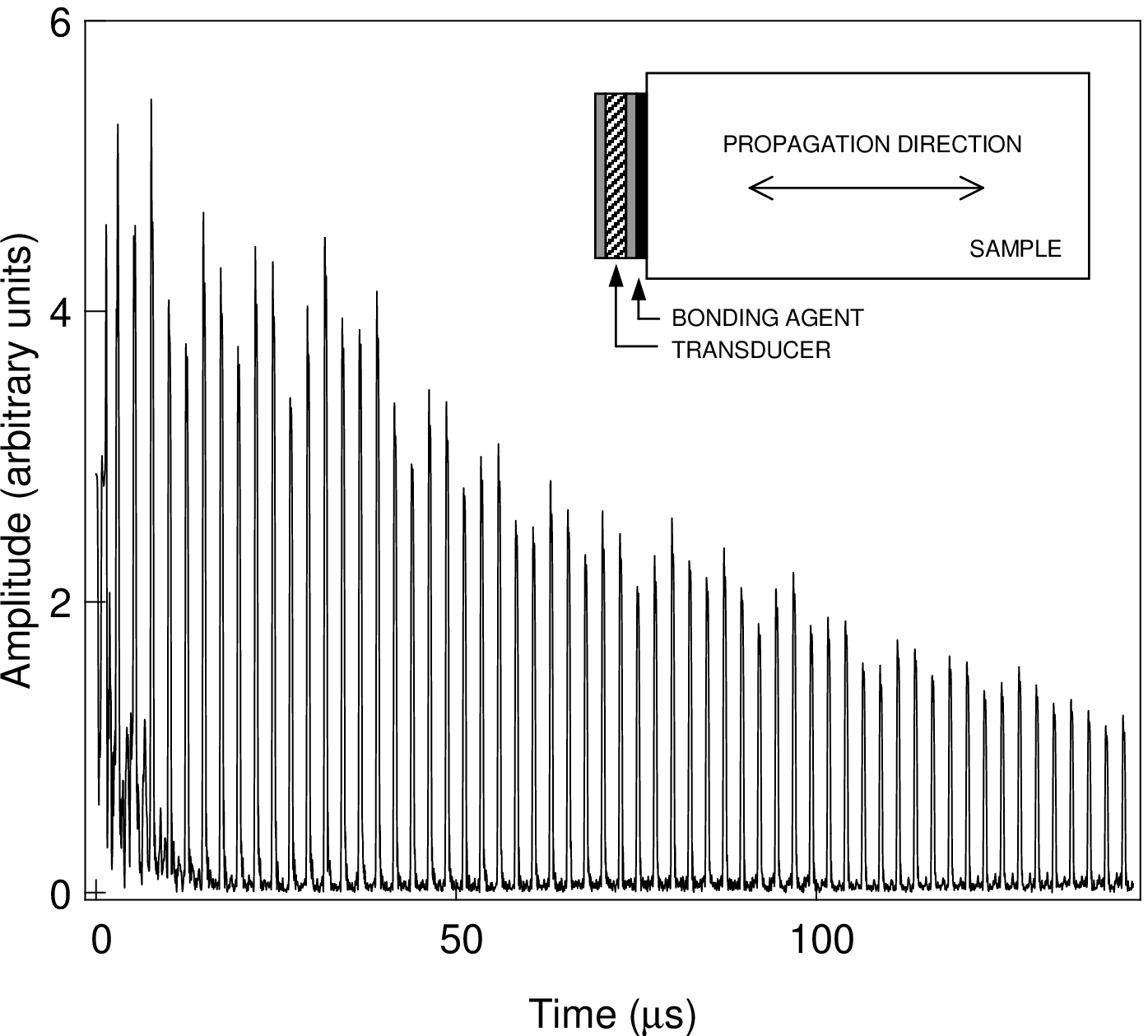}
 \caption{\label{fig:echoes} Echo pattern for a transverse sound mode corresponding to $C_{66}$, with 
propagation in the [100] crystal axis direction and polarization in the [010] direction of Sr$_2$RuO$_4$, 
obtained using the pulse-echo technique on a sample of length 3.98 mm (from Ref.~\protect\onlinecite{Lupien}). 
The inset shows a schematic of a typical pulse-echo setup.} \end{figure}

\begin{table*}
\caption{\label{tab:moduli} Single crystal elastic constants of Sr$_2$RuO$_4$. Units are in $10^{12}~$dynes/cm$^2$.}
\begin{ruledtabular}
\begin{tabular}{ccccccccc}
 Source & $C_{11}$ & $C_{33}$ & $C_{23}$ & $C_{12}$ & $C_{44}$ & $C_{66}$ & $(C_{11}-C_{12})/2$ & Temperature (K)\\
\hline
this work             &$2.32(2)$&$2.08(2)$&$0.71(2)$&$1.06(2)$&$0.657(4)$&$0.612(4)$&$0.63(1)$\footnotemark[1]&$300$\\
Ref.~\onlinecite{Lupien}&$2.35(7)$&          &       &$1.28(4)$&$0.68(2)$&$0.65(2)$ &$0.52(2)$ & $<4$\\
Ref.~\onlinecite{Lupien}\footnotemark[2]&$2.21(7)$&$2.4(2)$\footnotemark[3]  &       &$1.16(4)$&$0.66(2)$&$0.63(2)$ &$0.57(2)$ & $300$\\
Ref.~\onlinecite{Matsui67}&$\sim 1$ &$\sim 0.9$&       &         &$\sim 0.1$&$\sim 0.1$&          & $4.2$\\
Ref.~\onlinecite{Matsui69}&         &          &       &         &$0.4$     &$0.38$    &$0.37$    & $250$\\
Ref.~\onlinecite{Okuda} &$1.074$  &$0.783$   &$0.169$&$0.452$  &          &          &$0.311$   & $4.2$\\
Ref.~\onlinecite{Sarrao}~(La$_2$CuO$_4$) &$2.45$   &$2.48$    &$0.95$ &$0.61$   &$0.63$    &$0.52$     &$0.92$\footnotemark[1] & $580$\\
\end{tabular}
\end{ruledtabular}
\footnotetext[1]{Calculated for comparison using measured values of $C_{11}$ and $C_{12}$.}
\footnotetext[2]{$C_{ij}$ values extrapolated to 300 K using data from Ref.~\onlinecite{Matsui69}.}
\footnotetext[3]{Measured at 300 K.}
\end{table*}

The elastic constants reported by Lupien {\it et al.} were measured at low
temperature ($<4$~K) using the pulse-echo technique on samples cut from the same
ingot used in this study.\cite{Lupien} A typical pulse-echo experimental apparatus is
shown schematically in the inset of Fig.~\ref{fig:echoes}. With this technique, sound
velocities are calculated using the length between two parallel faces of an oriented
crystal and the measured time delay between echoes of the initial pulse, such as
those shown in Fig.~\ref{fig:echoes}.  By solving the usual Christoffel equations,
the elastic constants are obtained from sound velocities measured for various
propagation and polarization directions and the mass density.\cite{Truell} For
example, the shear elastic constant,
\begin{equation}
 C_{66} = \rho v_{[100/010]}^2,
\end{equation}
is calculated from the mass density $\rho$ and sound velocity $v_{[100/010]}$
(propagating in the [100] direction and polarized in the [010] direction). All
of values from Ref.~\onlinecite{Lupien} shown in Table~\ref{tab:moduli}, except
for $C_{12}$, are proportional to the square of various velocities directly
measured in different propagation and polarization directions (the value for
$C_{12}$ is given as an average of two values calculated using two different
sound velocities and the value of $C_{11}$).  By extrapolating these
low-temperature values to 300 K using temperature dependencies found
elsewhere,\cite{Matsui69} agreement is found to within experimental error for
$C_{44}$ and $C_{66}$, and to within $10\%$ for $C_{11}$, $C_{12}$ and $C_{33}$
when compared to the values in this study.  The various elastic constants
reported by Matsui {\it et al.}
(Ref.~\onlinecite{Matsui67},~\onlinecite{Matsui69}) and Okuda {\it et al.}
(Ref.~\onlinecite{Okuda}) are a factor of $\sim 2$ less than these values, which
is much greater than the $\sim 6\%$ change in the constants observed from low to
room temperature.

The Debye temperature $\theta_D$ is estimated from the elastic constants by using
the following expression for a tetragonal crystal,\cite{Alers}
\begin{equation}
 \theta_D = \frac{h}{k_B}\left(\frac{9N}{4\pi V}\right)^{-1/3}\rho^{-1/2}J^{-1/2},
\end{equation} 
where $h$ and $k_B$ are Planck's and Boltzmann's constants, respectively, $N/V$
is the atomic density, and $J$ is the series expansion of an integration over
sound velocities in all crystal directions, written in terms of the elastic
constants (see Eq.~32 in Ref.~\onlinecite{Alers}). Using Eq.~3, the full set of
elastic constants measured in this study gives $\theta_D = 465(5)$~K (this value
increases by $\sim 5$~K when using values extrapolated to low temperature in a
manner similar to that used in Table~\ref{tab:moduli} for
Ref.~\onlinecite{Lupien}). This elastic measure of $\theta_D$ compares well with
the value of $410(50)$~K extracted from the $T^3$ phonon contribution to
specific heat,\cite{thetaC} and the value of 427(50)~K obtained in M\"{o}ssbauer
measurements,\cite{thetaM} where $\theta_D$ is extracted from the temperature
dependence of the Debye-Waller factor.


In conclusion, the complete set of elastic constants was obtained for the tetragonal
crystal Sr$_2$RuO$_4$ using the measured resonance frequencies of a small rectagular
parallelepiped sample.  Excellent agreement was found with values obtained from
pulse-echo experiments performed on a sample cut from the same ingot.  The Debye
temperature determined from the measured constants was found to compare well with
values obtained from other measurements.


This work was supported by the Canadian Institute for Advanced Research and
funded by NSERC.  ~J.P. and C.L. acknowledge the support of the Walter C. Sumner
Foundation, and C.L.  also acknowledges the support of FCAR (Qu\'ebec).



\begin{references}


\bibitem{Maeno_Nature} Y. Maeno, H. Hashimoto, K. Yoshida, S. Nishizaki, T. Fujita, J. G. Bednorz and F. Lichtenberg, Nature (London) {\bf 372}, 532 (1994).

\bibitem{Maeno_PT} Y. Maeno, T. M. Rice and M. Sigrist, Phys. Today {\bf 54}, 42 (2001).

\bibitem{Truell} R. Truell, C. Elbaum and B.B. Chick, {\it Ultrasonic Methods in Solid State Physics} (Academic Press, New York, 1969), pp. 1-52.

\bibitem{Mao} Z. Q. Mao {\it et al.}, Mater. Res. Bull. {\bf 35}, 1813 (2000).

\bibitem{density} This value is obtained by calculating the density at 2~K using lattice constants measured via neutron powder diffraction \cite{Gardner}, and adjusting to room temperature for $0.6\%$ thermal expansion.\cite{Chmaissem}

\bibitem{Gardner} J. S. Gardner, G. Balakrishnan, D. McK. Paul and C. Haworth, Physica C {\bf 265}, 251 (1996).

\bibitem{Chmaissem} O. Chmaissem, J. D. Jorgensen, H. Shaked, S. Ikeda and Y. Maeno, Phys. Rev. B {\bf 57}, 5067 (1998).

\bibitem{Migliori} A. Migliori and J.L. Sarrao, {\it Resonant Ultrasound Spectroscopy} (Wiley, New York, 1997) and references therein.

\bibitem{JP} J. Paglione, M.Sc. Thesis (University of Toronto, 2000).

\bibitem{DeVisser} A. De Visser, A. Menovsky and J.J.M. Franse, Physica B {\bf 147}, 81 (1987).

\bibitem{Matsui67} H. Matsui, M. Yamaguchi, Y. Yoshida, A. Mukai, R. Settai, Y. \={O}nuki, H. Takei and N. Toyota, J. Phys. Soc. Jpn. {\bf 67}, 3687 (1998).

\bibitem{Sarrao} J.L. Sarrao, D. Mandrus, A. Migliori, Z. Fisk, I. Tanaka, H. Kojima, P.C. Canfield and P.D. Kodali, Phys. Rev. B {\bf 50}, 13 125 (1994).

\bibitem{error} The errors chosen according to dimensional bounds were obtained by uniformly increasing (decreasing) the three dimension input parameters in order to use the lower (upper) limit of the measured density in the fitting routine. It should be noted that any non-uniform change in the input dimensions, or in other words the shape of the sample, resulted in a fit with an extremely large RMS error between calculated and measured resonance frequencies, indicating the correct measurement of sample geometry.

\bibitem{Lupien} C. Lupien, W.A. MacFarlane, C. Proust, L. Taillefer, Z.Q. Mao and Y. Maeno, Phys. Rev. Lett. {\bf 86}, 5986 (2001);  C. Lupien, Ph.D. Thesis (University of Toronto, 2002).

\bibitem{Matsui69} H. Matsui, Y. Yoshida, A. Mukai, R. Settai, Y. \={O}nuki, H. Takei, N. Kimura, H. Aoki and N. Toyota, J. Phys. Soc. Jpn. {\bf 69}, 3769 (2000).

\bibitem{Okuda} N. Okuda, T. Suzuki, Z. Mao, Y. Maeno and T. Fujita, preprint submitted to SCES'2001 (2001).

\bibitem{Alers} G.A. Alers in {\it Physical Acoustics Vol. IIIB}, edited by W.P. Mason (Van Nostrand, Princeton, 1958).

\bibitem{thetaC} S. Nishizaki, Y. Maeno, S. Farner, S. Ikeda and T. Fujita, J. Phys. Soc. Jpn. {\bf 67}, 560 (1998).

\bibitem{thetaM} M. De Marco {\it et al.}, Phys. Rev. B {\bf 60}, 7570 (1999).


\end{references}
\end{document}